# Optical Probing of Ultrafast Laser-Induced Solid-to-Overdense-Plasma Transitions

Yasmina Azamoum[1, 2, *], Georg Alexander Becker[2], Sebastian Keppler[1, 2], Guillaume Duchateau[3], Stefan Skupin[4], Mickael Grech[5], Fabrice Catoire[6], Sebastian Hell[2], Issa Tamer[1,2], Marco Hornung[1, 2], Marco Hellwing[2], Alexander Kessler[1], Franck Schorcht[1], and Malte Christoph Kaluza[1, 2]

*Corresponding author: yasmina.azamoum@uni-jena.de

**Abstract**
Understanding the target dynamics during its interaction with a relativistic ultrashort laser pulse is a challenging fundamental multi-physics problem involving at least atomic and solid-state physics, plasma physics, and laser physics. Already, the properties of the so-called pre-plasma formed as the laser pulse's rising edge ionizes the target are complicated to access in experiments and modeling, and many aspects of this laser-induced transition from solid to overdense plasma over picosecond time scales are still open questions. At the same time, applications like laser-driven ion acceleration require precise knowledge and control of the pre-plasma because the efficiency of the acceleration process itself crucially depends on the target properties at the arrival of the relativistic intensity peak of the pulse. By capturing the dynamics of the initial stage of the interaction, we report on a detailed visualization of the pre-plasma formation and evolution. Nanometer-thin diamond-like carbon foils are shown to transition from solid to plasma during the laser rising edge with intensities $< 10^{16}$ W/cm². Single-shot near-infrared probe transmission measurements evidence sub-picosecond dynamics of an expanding plasma with densities above $10^{23}$ cm$^{-3}$ (about 100 times the critical plasma density). The complementarity of a solid-state interaction model and a kinetic plasma description provides deep insight into the interplay of ionization, collisions, and expansion.

**Introduction**

Since the turn of the millennium, the interaction of ultraintense laser pulses with thin foils has been shown to produce ion beams with unique properties[1,2], paving the way for groundbreaking applications like time-resolved radiography of electric and magnetic fields in plasmas[3], fast ignition in inertial confinement fusion[4,5], material testing and analysis[6], proton and carbon ion radiobiology[7,8] and cancer therapy[9]. The rapid progress in laser-driven ion acceleration has demonstrated the production of proton energies up to ~ 100 megaelectronvolts (MeV)[10], yet protons of ~ 200 MeV are required for radiation oncology[11]. Nevertheless, protons with energies of several hundred MeV were predicted theoretically[12,13], further motivating ongoing experimental endeavors[10,14].

Due to the complex nature of the interaction, several ion acceleration mechanisms can be triggered, which intricately depend on the laser pulse parameters, the properties of the target, the plasma, and their spatiotemporal evolution during the interaction. The laser pulse's temporal intensity profile, i.e., the laser contrast, exhibits a rising edge preceding the relativistic peak and may even include a pedestal due to amplified spontaneous emission (ASE) and ultra-short pre-pulses. Thus, in real-world experiments, the target is always ionized *before* the relativistic intensities are reached, forming a so-called pre-plasma. Efficient ion acceleration may occur depending on the pre-plasma state. For instance, in the case of Radiation Pressure Acceleration (RPA)[15] using nanometer-thin foils, one tries to minimize pre-plasma expansion at all costs because the strong pressure of the relativistic pulse is supposed to accelerate a thin layer of ions and electrons of a *non-expanded* target. Thus, RPA requires an ultrahigh laser contrast, which poses an extreme challenge for current state-of-the-art high-power lasers. In contrast, pre-plasma conditions can be tailored to optimize Target Normal Sheath Acceleration (TNSA)[16] employing µm-thick foils. The electrostatic field, which accelerates the protons at the target rear side, is induced by hot electrons heated by the main laser pulse in the pre-plasma produced on the target front side. Although extensively investigated, to date, proton energies achieved in TNSA[17] and RPA[18] have not yet exceeded 100 MeV.

Alternative acceleration schemes proposed in[12,13] are based on Relativistic-Induced Transparency (RIT)[19,20]. In contrast to surface acceleration as in TNSA and RPA, the laser peak penetrates a near-critical overdense pre-plasma due to the relativistic increase of the electrons' mass, which leads to a change in the plasma's refractive index (RIT effect). Hence, efficient heating of the particles may occur in the now extended interaction volume. Though appealing, the success of RIT-based schemes in experiments is challenging. Pre-expanded nanofoils are usually employed to this regime[10,14]. In this case, the pre-plasma density evolution must be tailored to the main pulse's rising edge.

The fine-tuning of the target state before the pulse peak's arrival requires accurate modeling to identify the tunable key parameters. In typical modeling approaches based on particle-in-cell (PIC) codes[21,22], the interaction is described starting from a pre-plasma state, which is then irradiated by the relativistic peak. The initial stage of the plasma formation from the solid state by the laser rising edge is usually ignored. However, several preponderant fundamental processes occur in this early interaction stage. Among others, these include initial ionization in the solid state, conduction electron heating by the laser, electron energy-coupling to the lattice or ions, phase transitions, and collisions. Simplifying the interplay of these processes is commonly done by making strong assumptions about the complex process of pre-plasma formation. Hydrodynamic codes[23,24] are usually employed to infer pre-plasma properties considering the laser's rising edge and possible pre-pulses. Although these codes provide a reasonable estimate of the spatial density profiles, the distribution functions of the species, i.e., their temperatures and the ionization state, are based on approximations. For instance, only averaged physical quantities (densities, velocities, etc.) are considered in these codes, assuming the distribution functions of particles at equilibrium. This oversimplification may result in an inaccurate prediction of the pre-plasma properties and the acceleration process, motivating further experimental and theoretical investigations.

In experiments, capturing the ultrafast evolution of the pre-plasma during the steep laser rising edge, i.e., femtosecond (fs) dynamics on the nanometer (nm) scale, is challenging. Moreover, the shot-to-shot fluctuations inherent to high-power lasers make single-shot probing diagnostics desirable.

Optical probing is a convenient tool to investigate such plasmas. Light with wavelength $\lambda$ can only propagate in a plasma with electron density $n_e < n_c$ where $n_c$ is the critical density given by $n_c = \omega^2 m_e \varepsilon_0 / e^2 \approx 1.11 \times 10^{21}$ cm$^{-3}/[\lambda(\mu m)]^2$. Then, a non-collisional, non-magnetized, and non-relativistic plasma exhibits a real-valued refractive index $\eta = \sqrt{1 - n_e/n_c}$. The nm-scale pre-plasma expansion was recently measured with reflected visible probe light from the target[25]. In contrast, the density and the temperature evolutions of plasmas which are overdense for near-infrared (NIR) light, $n_e > 10^{21}$ cm$^{-3}$, could only be diagnosed using laser-driven XUV or hard X-rays sources in combination with hydrodynamic codes or X-ray absorption spectroscopy with computationally demanding ab initio calculations[26,27] (sub-ps time resolution), or using x-ray free electron lasers[28] (fs- and nm- resolutions). While these techniques give access to overdense plasma properties and an accurate description at the atomic level, they imply using limited-access facilities or, at least, a rather complex setup for the probe. For none of the mentioned methods, single-shot probing was reported.

In this paper, we will exploit that longer probe wavelengths, e.g., in the near-infrared (NIR) regime, can still be used to diagnose such plasmas. When $n_e > n_c$, the probe light is primarily reflected as $\eta$ becomes imaginary. However, a fraction of the light still penetrates the plasma over the skin depth $l_s \approx c/\omega_p$, where $\omega_p = \sqrt{n_e e^2/(m_e \varepsilon_0)}$ is the plasma frequency. If the plasma is sufficiently thin ($< l_s$), the NIR light tunnels through and can be detected to investigate the target's dynamics. Besides, as $l_s \propto 1/\sqrt{n_e}$, a large electron density range can be probed in ultrathin targets. Furthermore, when the pulse is temporally chirped, the probe light allows the investigation of time-dependent plasma dynamics in single-shot measurements[29].

To model such optical probing of thin, overdense plasma dynamics, we propose a novel and alternative approach to describe, with a limited number of assumptions, the pre-plasma formation from the initial laser-target interaction, namely, the transition from the solid state to the plasma state. The further pre-plasma evolution before the arrival of the main pulse peak can be readily described using well-established PIC codes. So far, such a transition has only been investigated at lower peak intensities ($< 10^{13}$ W/cm$^2$), using laser pulses ($\sim 100$ ps) with reported plasma dynamics on tens of picoseconds (ps) time scales[30]. In contrast, an ultrafast transition was only investigated in experiments with $< 100$ fs-resolution[31] using sub-picosecond laser pulses. However, to our knowledge, the ultrafast solid-to-plasma transition during a steep laser rising edge has not yet been described in detail.

Using a pump-probe approach comprising single-shot NIR probe light transmission measurements and a two-step interaction model, we report on the experimental observation of an ultrafast transition from solid to highly overdense and expanding plasma. The latter is triggered during the laser rising edge when irradiating nm-thin diamond-like carbon (DLC) foils by femtosecond laser pulses with peak intensities exceeding I $\sim 10^{16}$ W/cm$^2$. Similar parameters can readily be employed when using a controlled pre-pulse interacting with a relativistic peak in the context of RIT, which additionally motivates this study.

**Results**

The evolution of the laser contrast of the pump pulses, used to irradiate nm-thick DLC foils, is depicted in Fig. 1a. The profile is well described by the two fitting curves in red, which are used in the upcoming data analysis and modeling. Particularly relevant will be the steep rising edge in the time window $-3.7$ ps $\leq t_{\text{pump}} \leq -0.2$ ps, described by the contrast ratio CR $= \exp(-|t_{\text{pump}}|/277$ fs$)$ with an intensity profile $I = I_{\text{peak}} \times$ CR, where $I_{\text{peak}}$ is the maximum intensity at the pulse's peak. To ensure that the plasma is formed during this steep rising edge only (around an intensity $I \sim 10^{12}$ W/cm$^2$), $I_{\text{peak}}$ is reduced to $\sim 10^{15}$ W/cm$^2$ so that significant ionization starts after $t_{\text{pump}} \sim -4$ ps. As the steep rising edge covers $\sim 5$ orders of magnitude in intensity, varying the peak intensity by one order of magnitude may result in a relative time shift of the plasma formation. However, due to the particular functional form of the contrast ratio CR, besides this shift, the temporal evolution of the plasma remains unchanged. The experimental setup is shown in Fig. 1b. The interaction region was diagnosed in the transverse $y$-direction and in time, using a temporally-chirped broadband probe pulse whose different wavelengths arrive at different interaction times. With this approach, the plasma formation and evolution can be recorded within a single shot, achieving a sub-ps time resolution[32] (see details for pulse characterization and probing in Methods).

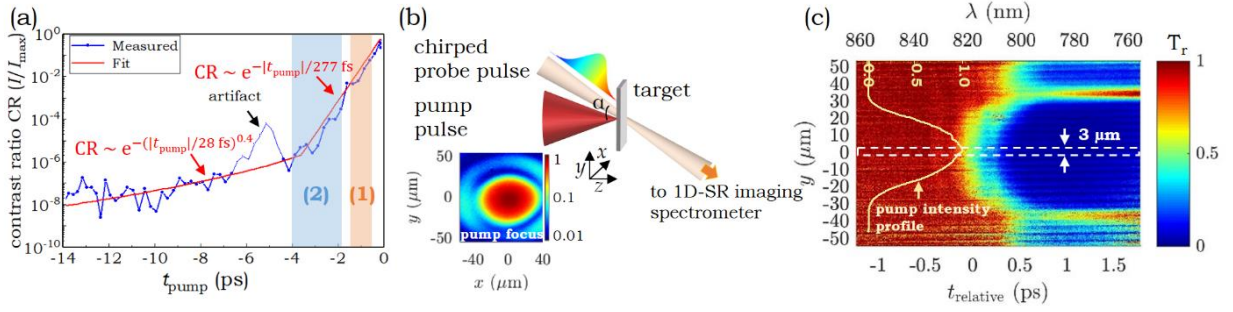

**Fig. 1 :** Single-shot space and time-resolved probe transmission measurement. (a) The pump laser's temporal intensity contrast. The shaded regions (1) and (2) are related to the modeling discussed in Fig. 3. The feature at $t_{\text{pump}} \sim -5$ ps is an artifact from the measurement, cf. Methods. (b) The experimental arrangement. The pump pulse is at normal incidence on the target, while the probe pulse is obliquely incident under an angle $\alpha = 37°$. The inset shows the pump focus's normalized spatial intensity profile. (c) 1D-spatially and temporally resolved relative transmission $T_r$ of the probe for a 10 nm-thick DLC foil measured using a 1D-spatially resolving (SR) imaging spectrometer and a chirped probe pulse. The wavelength on the top axis is converted into a relative time on the bottom axis. The inserted yellow curve is the normalized spatial intensity profile of the pump pulse in the transverse y-direction.

A typical space- and time-resolved probe transmission map measured with a 1D spatially-resolved imaging spectrometer through a main-pulse irradiated, 10-nm thick DLC foil as the target is shown in Fig. 1c. The map reveals the transition from the target being transparent to an opaque state where the probe is blocked for $\lambda \lesssim 820$ nm and hence $T_r \sim 0$. The plasma profile in the transverse y-direction exhibits a shape and size similar to the focal spot, cf. inset of Fig. 1b, indicating that the laser's first low-intensity Airy ring beyond the first minimum induces these wings.

In the upcoming analysis we focus on the plasma dynamics in the central high-intensity region of the focal spot at $y = 0$ μm. For the sake of simplicity, throughout the paper the particle densities are expressed as a function of the critical plasma density $n_c = 1.72 \times 10^{21}$ cm$^{-3}$ for the probe wavelength $\lambda = 800$ nm. We observed in our simulations that the plasma dispersion is negligible in the probe wavelength range $\lambda \approx 700 - 900$ nm. DLC foils of various thicknesses (5, 10, 20, 50 nm) were used.

The measured absolute transmissions $T$ for all DLC thicknesses are depicted as the blue lines in Fig. 2 (see Methods for measurements and processing details). The transmission profiles for each foil are reproducible when varying the peak intensity over one order of magnitude, which contrasts with previous works[30,31]. This result confirms that the plasma always forms during the steep rising edge described by $\text{CR} = \exp(-|t_{\text{pump}}|/277 \text{ fs})$. Varying the peak intensity only shifts the plasma formation in time, i.e., by $\sim 1$ ps, but the temporal evolution remains the same. Therefore, the knowledge of the absolute timing of the probe's arrival is not required in the framework of this investigation. The time axes in Fig. 2 are, hence, expressed in relative times where $t_{\text{relative}} = 0$ ps is set to the value $T_r \sim 50$ % corresponding to the inflection point of the transmission profile. The profiles show a sub-picosecond transition from a transparent, solid-state target foil represented by the plateau ($T = T_0$) at early times to an overdense plasma state at later times ($T \sim 0$). The transmission dynamics can be characterized by the thickness-dependent time $\tau \sim 400 - 700$ fs required for the transmission to drop from 90 % to 10 %.

As the targets are very thin (thickness $\ll \lambda$), the low measured probe transmissions ($T \sim 0.01$) imply highly overdense plasmas with electron densities $n_e \gg n_c$. To evaluate the impact of the plasma electron density on the probe transmission, we calculate the optical tunneling of the probe through a homogeneous plasma slab analytically (see Methods.). For a plasma slab of 10 nm-thickness with electron density $n_e = 50 n_c$, this estimation leads to $T \sim 0.26$. Hence, a significant fraction of the probe intensity is expected to tunnel through the target. However, our measurements yield even lower transmission values, indicating that the plasma is highly overdense $> 50 n_c$. Furthermore, assuming a plasma slab of 5 nm thickness with electron density $n_e^{\text{fi}} = 371 n_c$, corresponding to full ionization of the DLC foil (see Methods for target properties), the previous plasma slab model would give $T \sim 0.024$, which is significantly higher than the measured value of 0.01. For comparison, a density of $\sim 2 n_e^{\text{fi}}$ would decrease the transmission to 0.5 %. Since the plasma density cannot exceed $n_e^{\text{fi}}$, yet another process must be responsible for the reduction of probe light transmission in the experiment. The simple plasma slab model shows that for $n_e \gg n_c$, increasing the plasma thickness $d$ while keeping the product $n_e d$ constant decreases the transmission. Therefore, our measurements confirm that significant plasma expansion already occurs on sub-ps time-scales, and thus needs to be considered in the modeling.

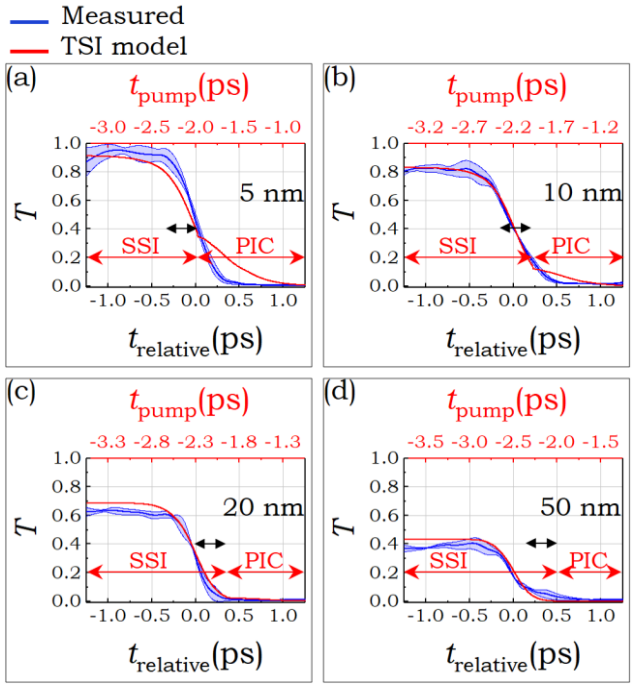

**Fig. 2**: Measured (blue) and calculated (red) absolute transmission $T(t)$ for DLC foils of thicknesses from 5 to 50 nm. The measurements are averaged over four shots with peak intensities $I_{\text{peak}} \sim 10^{15} - 10^{16}$ W/cm². The shaded region is the standard deviation over all shots for each foil. The red curve is computed using the TSI model. The measured and calculated curves are superimposed at their inflection points at $t_{\text{relative}} = 0$ ps, corresponding to 50 % relative transmission. The red and the black double arrows delimit the time intervals for SSI and PIC and the extended region of the SSI to reach the melted state, respectively.

**Discussion**

Comprehensive numerical simulations were carried out to explain the experimental findings. We used two complementary interaction models to compute the time-dependent free electron density $n_e(t)$. As the transmission profiles in the experiment were found to be insensitive to the peak intensity and the plasma forms during the steep rising edge of the laser pulse, the laser intensity is described by $I = I_0 \exp(-|t_{\text{pump}}|/277 \text{ fs})$ in the simulations with $I_0 = 10^{15}$ W/cm². The transmission of a probe plane wave propagating through the generated plasma with density $n_e(z, t)$ was calculated to compare the simulation results to the experimental measurements. Further details of our computational methodology are provided in Methods.

In a first attempt, the interaction was simulated using the one-dimensional (1D) PIC code SMILEI[22], considering a cold DLC foil interacting with the pump pulse (simulation parameters are given in Methods). Fig. 3a shows the resulting transmission profiles using the electron densities obtained from the SMILEI PIC code alone. For all target thicknesses, the transmission drops quickly with $\tau \lesssim 100$ fs, i.e. much faster than experimentally measured. In addition, Fig. 3b shows that the plasma formation (ionization) starts at $t_{\text{pump}} \sim -1$ ps where $I \sim 10^{13}$ W/cm², which is the intensity threshold for photo-ionization of carbon atoms. This ultrafast transition from a cold target to a highly overdense plasma, i.e., $> 100 n_c$ over a time interval $< 100$ fs, corresponds to the abrupt drop of the predicted probe transmission.

This discrepancy with the experiment can be attributed to the inadequate description of the pristine target. The PIC code treats the target as an ensemble of individual carbon atoms, not accounting for the solid state. Hence, only the ionization of the atoms is considered. However, DLC may already be ionized as a solid carbon foil at lower energy. Indeed, the target is a semiconductor with a band gap of ~ 1.1 eV. This energy value is much smaller than the first ionization energy of carbon atoms ~ 11.3 eV. The plasma is, thus, expected to be formed earlier and at lower laser intensities. Therefore, the actual ionization dynamics of the solid foil need to be included in the modeling.

To correctly account for ionization in solids, a simulation was carried out with a solid-state interaction (SSI) model adapted from Ref.[33]. The ionization is described by solving multiple rate equations[34] (details are given in Methods). This model was already used to successfully interpret an experiment of laser-induced plasma formation from a dielectric solid presented in Ref.[30]. The SSI simulations take into account the full pump laser's temporal intensity profile (Fig. 1a, solid red line), yielding the temporal evolution of the plasma density $n_e(t)$ as shown by the blue line in Fig. 4a.

As expected, significant plasma generation occurs in the steep rising edge $t_{\text{pump}} > -3$ ps. By assuming homogeneous, non-expanding plasmas with thicknesses corresponding to the target foils, the transmission profiles can be computed and are shown in Fig. 3c. The SSI model stops at $t_{\text{pump}} \sim -2$ ps in the rising edge where the maximum density $n_e \sim 70 n_c$ is achieved. The original SSI model was developed for dielectric materials such as fused silica $SiO_2$ with a high band gap ~ 9 eV and a maximum density $n_e \sim 20 n_c$. Because DLC is a semiconductor with a much lower band gap (~ 1.1 eV), further ionization of inner shells in the band structure may occur. Thus, higher electron density may be produced, and the model validity has been reasonably extended to $n_e \sim 70 n_c$ for our configuration (this value will be further discussed below). Therefore, such highly overdense plasma may be described by the model. However, going beyond this value would certainly break the model's validity, e.g., because of the ionization of the inner shells of carbon atoms. This process is not adequately accounted for in the SSI model. Therefore, except for the 50 nm foil, the transmission dynamics are not fully described by this model alone, either, since total opaqueness $T \sim 0$ is not reached at $t_{\text{pump}} \sim -2$ ps. Nevertheless, we observe a significantly slower transmission decrease in the SSI model compared to the previous PIC results in Fig. 3a, closer to our experimental observations. Furthermore, the density evolution in Fig. 3d indicates that ionization starts earlier and, thus, at lower intensities than predicted by the PIC code alone involving atomic ionization rates.

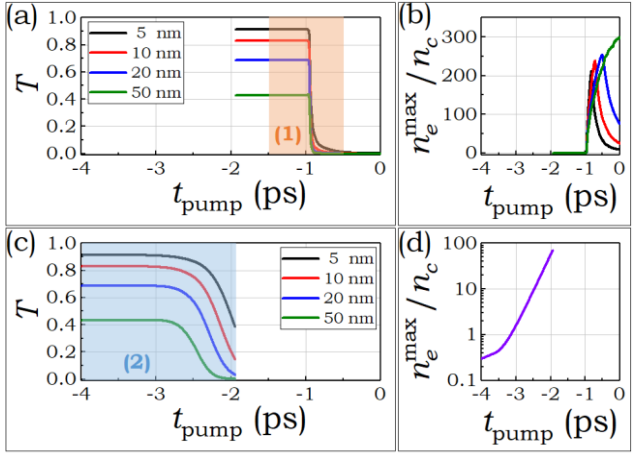

**Fig. 3**: Computed time-dependent probe transmission and the corresponding electron densities for initially cold DLC foils with thicknesses ranging from 5 to 50 nm interacting with the pump laser using different models. (a) and (b) show the $T(t)$ and the maximum density $n_e^{\max}$ along the z-axis as a function of time, respectively computed using the SMILEI PIC code. (c) and (d) give the $T(t)$ and $n_e^{\max}$, respectively, computed using the SSI model where $n_e^{\max} = n_e$, since the plasma is assumed to be spatially homogeneous (extracted from Fig. 4a). The SSI model validity stops at $n_e \sim 70 n_c$ at $I \sim 10^{12}$ W/cm² see details in the text. The shaded regions (1) and (2) correspond to time intervals indicated in the laser temporal intensity contrast in Fig. 1(a).

**Solid-state and kinetic plasma description: Two-step model**

To overcome the limitations of both models and provide a better description of the target dynamics, we propose a combination of the SSI model at earlier times and the PIC description at later times. On the one hand, the SSI model describes the laser interaction with solids well, including the initial ionization. On the other hand, the PIC code correctly handles the kinetics of a highly overdense plasma, including plasma expansion and inner shells' ionization process. We will refer to the combination of SSI and PIC as the two-step interaction (TSI) model. In the TSI model, the simulation starts with the SSI model and is continued by a PIC simulation after an overdense plasma is formed. As the PIC description considers only free particles like electrons, ions, and atoms, a reasonable switching point is when the melting state of DLC is reached. At this point, the band structure disappears, and the ions start to be free.

A semiconductor under femtosecond laser excitation may undergo non-thermal melting[35,36]. In contrast to thermal melting, a significant fraction of the electrons is promoted abruptly from the valence band to the conduction band. Consequently, the lattice bonds are rapidly weakened, and the ions or atoms start to move before reaching the thermal melting point. This scenario may occur as an exponentially increasing laser intensity continuously irradiates the DLC. The non-thermal melting may start at $n_e \sim 10 n_c$ [35], the threshold for Si, a semiconductor with a band gap of $\sim$ 1.12 eV similar to DLC. Additionally, ions require a few 100 fs to be entirely free[35]. Therefore, to ensure an initial plasma state composed of entirely free ions as assumed in the PIC description, we extrapolate the SSI model to $n_e^m \sim 70 n_c$ so that the PIC simulation starts about 0.5 ps after the beginning of the melting process. To bridge the SSI ionization dynamics to the PIC simulations when non-thermal melting occurs, electron and lattice temperatures, $T_e$ and $T_l$, respectively, must be determined. Following Refs. [30,33] and references therein, we use a standard two-temperature model (TTM). The results shown in Fig. 4a indicate that, at the melting, $n_e^m$ is reached at $t_{\text{pump}} \sim -1.93$ ps, corresponding to an intensity of $I \sim 10^{12}$ W/cm² and $T_e^m \sim 4.6$ eV and $T_i^m \sim 0.34$ eV. (Further details on the TSI parameters are provided in Methods.).

The calculated transmission dynamics using the densities $n_e(z,t)$ from the TSI model are shown in Fig. 2. All transmission profiles reach $T \sim 0$ in the PIC stage of the model, where plasma expansion is considered. The characteristic times $\tau$ of the transmission dynamics are a few hundreds of fs, which agrees well with the measurements. The calculated transmission curves are superimposed with their experimental counterparts for each foil thickness by overlapping their respective inflection points at $t_{\text{relative}} = 0$ ps. It is worth noting that extrapolating the SSI model to $n_e \sim 70 n_c$ shows an excellent agreement with the experiment, further validating the SSI model's application to lower band gap materials such as DLC. The extended SSI domain is indicated by the black double arrows in Fig. 2. Some slight discrepancies between the TSI results and our measurements can be observed, such as a transmission shift in the plateau region (5, 20, and 50 nm cases), a higher transmission predicted by the TSI model for the PIC results with the thinnest foils, and the reverse behavior for the 50 nm case. Possible reasons may be the experimental target thickness uncertainty of about 20 % or a systematic underestimation of the ionization yield and thus $n_e$ in the PIC simulations due to the abrupt switch to a pure particle description. Besides, an overestimation of $n_e$ in the SSI description can also be attributed to an inhomogeneous ionization likely occurring for thicker foils because $l_s$ is shorter than the plasma thickness. Such target inhomogeneity is currently not accounted for in the first part of the TSI model. Moreover, during its transition from solid to plasma, the target passes through the highly non-linear and -- for our conditions -- ultrafast regime of warm dense matter (WDM), which neither SSI nor PIC models adequately describe. Considering and mitigating these limitations is beyond the scope of this work. Nevertheless, our TSI approach confirms that an ultrafast solid-to-overdense plasma occurs in the experiments. To correctly describe this early stage of the interaction, both the solid (SSI model) and the plasma (PIC approach) properties are essential, and only the combination of both models yields an accurate description of the experimental measurements.

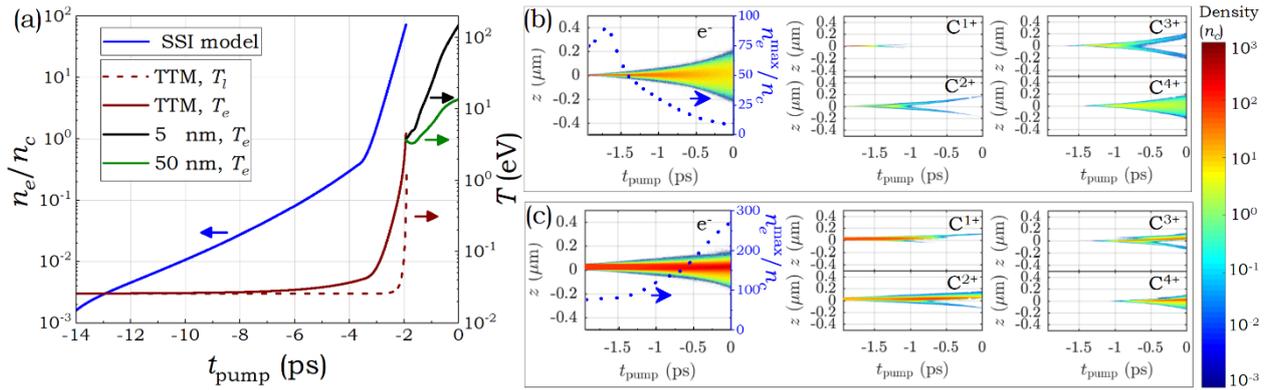

**Fig. 4**: Computed plasma properties during the interaction from the TSI model for DLC foils of 5 and 50 nm thicknesses. Here, the pump laser propagates in positive z-direction. (a) Time-dependent electron density from the SSI model, lattice ($T_l$), and electron ($T_e$) temperatures from the TTM model for all thicknesses and $T_e$ for 5 and 50 nm thicknesses from PIC simulations. (b) and (c) show the spatiotemporal dynamics of the electron and the carbon ion densities for 5 and 50 nm thick foils, respectively. The blue dotted lines correspond to the maximum electron density on the z-axis as a function of time.

**Initial plasma expansion and interplay of ionization processes**

To emphasize the role of the plasma expansion, Fig. 4b,c shows the spatiotemporal evolution of the plasma properties for the 5 nm and the 50 nm-thick foils. For the thinnest foil in Fig. 4b, $n_e(z,t)$ exhibits a strongly expanded profile at the end of the simulation, caused by the rapid heating of the plasma as the intensity approaches the peak. At $t_{\text{pump}} = 0$ ps, the plasma thickness with density $n_e \geq n_c$ is estimated to be ~ 300 nm, about 60 times the original target thickness. Besides, the relatively low maximum density of $n_e \sim 7n_c$ highlights the importance of the plasma expansion for lowering the probe transmission. Additionally, the density profile evolves symmetrically in time, in contrast to the strong asymmetry observed for the 50 nm-thick foil shown in Fig. 4c, where a high-density region reaching a maximum of ~ $270n_c$ with extended low-density drops on the target front and back side, while the drop is steeper at the back. These results point out that, on the one hand, our measurements evidenced the plasma expansion for the thinnest foil, and, on the other hand, applying the TSI model is crucial for correctly describing the interaction in its initial phase. Such detailed knowledge of the target evolution is paramount to match the laser contrast conditions to the target (or plasma) thickness to achieve efficient laser-driven ion acceleration.

Finally, the interplay of fundamental processes such as ionization and collisions, which eventually determine the target properties at the peak arrival, are accessible using our experimental measurements and their comparison to our modeling strategy. During the laser-induced solid-to-plasma transition investigated in this work, the free electrons are produced in the SSI step by the MPI process in the solid state. Due to the low electron temperature during this step ($T_e < 5$ eV) inferred from the TTM, cf. Fig. 4a, collisional ionization (CI) is negligible, estimated to be ~ 1 %. In fact, although $n_e \sim 70n_c$ at $t_{\text{pump}} \sim -2$ ps, the laser intensity $I \sim 10^{12}$ W/cm$^2$ is not sufficient to heat the electrons to induce a significant number of ionizing collisions. In the PIC step, however, collisions occur as free electrons are available and heated by the exponentially increasing laser intensity. Consequently, CI starts at a lower intensity than the threshold intensity $I \sim 10^{13}$ W/cm$^2$ for photo-ionization of carbon ions and becomes quickly dominant. The abrupt behavior of CI, being negligible at the end of SSI and being dominant at the beginning of the PIC phase, suggests that CI starts to become dominant during the highly non-linear and ultrafast WDM transition, which our TSI model does not describe. Therefore, despite being in an intensity range suitable for photo-ionizing carbon ions, this ionization process does not play a role in our experiments. Simulations without photo-ionization in SMILEI show the same results as in Fig. 2 and thus confirm this interpretation. While the ionization charge state of the pre-plasma is usually based on assumptions when modeling relativistic laser-matter interaction, in this work, gaining insight into the ionization processes leads naturally to the detailed knowledge of different ion species and their dynamics in the plasma. For example, the final average charge state $C^{4+}$ shown in Fig. 4b,c could not have been predicted without considering collisions and CI being the dominant ionization mechanism.

**Conclusions**

In summary, our investigation sheds light on the sub-picosecond transition from a solid target to a highly overdense plasma ($n_e > 100 n_c$) produced with nm-thin DLC foils during the laser rising edge with intensities increasing up to $I \sim 10^{16}$ W/cm$^2$. Even though this stage of pre-plasma formation is crucial for the conditions for the subsequent ion acceleration during a relativistic laser-thin foil interaction, this transition has neither been studied in detail in simulations nor detected in experiments. We demonstrated an all-optical single-shot technique that characterizes the complete target evolution. Because our technique relies on optical tunneling, accessing the overdense plasma regime is possible. Our findings indicate that correctly describing the target transition from a solid to a plasma state is crucial for understanding the plasma evolution in such laser-solid interaction. Our single-shot NIR probe transmission measurements evidence a non-negligible plasma expansion that can significantly reduce the probe intensity tunneling through very thin foils. We develop a general picture of the evolution of the plasma by employing a two-step interaction model comprising a combination of a solid-state interaction model and a PIC code. A detailed description of the pre-plasma properties before the peak arrival is achieved, going well beyond the previous modeling of relativistic laser-matter interactions. Our approach can readily provide a detailed description of the plasma formed by pre-expanding a thin foil using a controlled pre-pulse, usually in the intensity range studied in this work. Besides being of fundamental interest, such insight is crucial to finding the matching laser-target conditions required for the RIT-based acceleration regime. Our experimental findings and the application of our modeling approaches might therefore help to bring laser-accelerated ion technologies to societal applications.

**Methods**

**Laser system and pulse characterization**

The experiments were carried out using the all-diode pumped high-power laser system POLARIS[37], operated by the Helmholtz-Institute Jena and the Institute of Optics and Quantum Electronics in Jena. The temporal intensity contrast of the pump laser, as measured with a third-order cross-correlator (Amplitude, Sequoia), is shown in Fig. 1a. The profile is well described by the two fitting curves in red. The indicated artifact at $t_{\text{pump}} \sim -5$ ps is due to a post-pulse induced by a glass wafer inserted in the beam path for debris shielding, therefore not affecting the interaction, and is therefore ignored throughout our analysis.

Particularly relevant is the steep rising edge in the time window $-3.7$ ps $\leq t_{\text{pump}} \leq -0.2$ ps, described by the contrast ratio CR = $\exp(-|t_{\text{pump}}|/277$ fs$)$ with an intensity profile $I = I_{\text{peak}} \times$ CR, where $I_{\text{peak}}$ is the intensity of the peak characterized by $\sim 150$ fs Full Width at Half Maximum (FWHM) pulse duration. In the experiments, the peak intensity is reduced to $I_{\text{peak}} \sim 10^{15}$ W/cm$^2$ by inserting a half-inch aperture in the beam path of the few J-energy, 140 mm-diameter, and linearly polarized pulses centered at $\lambda_p = 1030$ nm before being focused with an off-axis parabola (300 mm-focal length) at normal incidence on the DLC foil. Thus, the relative contrast profile is kept similar to that for the relativistic pulses. The resulting $\approx 3$ mJ energy pump pulses are focused to a spot showing an Airy pattern (see inset in Fig. 1b) with a $\approx 40$ µm FWHM diameter containing $\sim 60$ % of the pulse energy after the aperture.

**Target**

The diamond-like carbon (DLC) targets used in this experiment are free-standing foils of pure carbon, produced by pulsed laser deposition technique[38] with a mass density of $\rho_{\text{DLC}} = 2.15$ g/cm$^3$. The DLC foil is an amorphous semiconductor[39], characterized by an electronic band structure with a band gap of $\sim 1.1$ eV. The latter was estimated using the Tauc method[40]. The target refractive index is given by $\eta_{\text{DLC}} = n_{\text{DLC}} + i\kappa_{\text{DLC}}$ where $n_{\text{DLC}} \approx 2.65$[41] and the extinction coefficient $\kappa_{\text{DLC}} \approx 0.5$ were obtained from a wavelength-dependent transmission measurement carried out using a Shimadzu Solid Spec-3700-spectrometer. The carbon ionization energies are 11.3, 24.4, 47.8, 64.5, 392, and 490 eV.

**Single-shot space and time-resolved probe transmission diagnostic**

The plasma dynamics are investigated by longitudinally irradiating a $\sim 100$ µm-extended region of the interaction with $p$-polarized broadband ($\Delta\lambda \approx 150$ nm centered at $\lambda \approx 840$ nm under an incidence angle of $\alpha = 37°$) and $\approx 12$ µJ-energy probe pulses produced in a Non-collinear Optical Parametric Amplifier (NOPA)[42]. When optimally compressed, the probe pulses have a duration of $\approx 14$ fs. However, by applying a positive chirp to the pulses, their duration is stretched to $\sim 6$ ps so that their different wavelength components arrive at different interaction times. With this approach, the plasma formation and evolution can be recorded within a single shot, achieving a sub-ps time resolution[32]. An extent of $\Delta x \approx 3$ µm of the interaction region is imaged onto the entrance slit of the 1D-spatially resolving spectrometer. The relative timing between POLARIS-main and NOPA probe pulses (seeded by the same oscillator) can be adjusted using a delay stage with sub-ps resolution. We measure the relative probe transmission through the plasma $T_r = T/T_0$, where $T_0$ and $T$ are the measured transmission values without and with the interaction induced by the pump pulse, respectively. The conversion of the probe wavelength to time is calibrated using an additional pre-pulse with an adjustable time delay. The measured absolute transmission $T$ in Fig. 2 is obtained as the lineout at $y = 0$ μm averaged by 3 μm. The measurements in the same figure are averaged over four shots with intensities in the range $I \sim 10^{15} - 10^{16}$ W/cm² for each foil thickness.

**Modeling**
**Analytical model for optical tunneling**

We assume a $p$-polarized probe plane wave at an angle of incidence of $\alpha = 37°$ at the overdense and homogeneous plasma slab of thickness $d$ with the dielectric function $\varepsilon < 0$. Then, exploiting Maxwell's boundary conditions[43] at the front and rear side of the slab yields the transmission

$$T = \frac{4e^{-4\pi\gamma d/\lambda}}{(1+C^2)(1+e^{-8\pi\gamma d/\lambda}) + 2(1-C^2)e^{-4\pi\gamma d/\lambda}}$$

with $\gamma = \sqrt{\sin^2\alpha - \varepsilon}$, $C = \frac{\gamma^2 - \varepsilon^2\cos^2\alpha}{2\varepsilon\gamma\cos\alpha}$ and $\lambda$ being the central probe wavelength. For a highly overdense plasma, we can assume $\varepsilon \approx 1 - n_e/n_c$.

**Numerical simulations**

We used two complementary interaction models, discussed in detail below, to compute the time-dependent free electron density $n_e(t)$. As the transmission profiles in the experiment were found to be insensitive to the peak intensity and the plasma forms during the steep rising edge of the laser pulse, the laser intensity is described by $I = I_0 \exp(-|t_{\text{pump}}|/277 \text{ fs})$ in the simulations, and $I_0 = 10^{15}$ W/cm². The transmission of a probe plane wave propagating through the generated plasma with density $n_e(z,t)$ was calculated by solving Maxwell's equations. To this end, the matrix method presented, e.g., in Refs. [44,45] was adapted. The Drude Model is used to compute the complex dielectric function $\varepsilon$ expressed as a function of $n_e$ as

$$\varepsilon = \eta_{\text{DLC}}^2 - \frac{n_e}{n_c}\left(1 + i\frac{\nu_c}{\omega}\right)$$

with $\eta_{\text{DLC}}$ being the refractive index of the pristine target (see target section.); the temperature-averaged electron-ion collision frequency was chosen as $\nu_c = 5 \times 10^{14}$ s⁻¹, consistent with what is discussed and used in the two-temperature model; $\omega$ is the probe's angular frequency.

**a. Particle-in-cell simulation (PIC)**

We use the one-dimensional (1D) PIC code SMILEI[22]. The relevant implemented processes include multiphoton ionization (MPI), field ionization (tunneling ionization, TI) and collisional ionization (CI) for atoms, and binary collisions between electrons and ions. In the simulation box of $L = 1$ μm length with a cell size of $\Delta z = 0.156$ nm, the target was modeled as a slab of cold carbon atoms at solid density $n_a = 62n_c$, positioned at $z = 0$ μm (cf. Fig. 4b, c). The pump laser with a central wavelength $\lambda_p = 1030$ nm enters the box from $z < 0$. Its temporal intensity envelope follows the steep rising edge mentioned above. The temporal resolution was $\Delta t = 5 \times 10^{-4}$ fs. The number of particles per cell was initialized as follows: 2000 carbon atoms per cell were used in the simulation using cold target, and 1452, 274, and 2000 particles per cell were used in the TSI model for the carbon ions $C^{1+}$, $C^{2+}$ and electrons, respectively.

**b. Solid-state interaction model (SSI)**

To correctly account for ionization in solids, we use a solid-state interaction (SSI) model adapted from Ref.[33]. In this model, the ionization is described by solving state-of-the-art multiple rate equations[34] where the target band structure is described by a set of states accounting for the electron dynamics in the conduction band. An electron density $n_i$ is associated with each state (where $i \in 0, 1, 2$), and the coupled system reads

$$\frac{\partial n_0}{\partial t} = W_{\text{PI}} + 2\tilde{\alpha}n_2 - W_1 n_0 - n_0/\tau_r,$$

$$\frac{\partial n_1}{\partial t} = W_1 n_0 - W_1 n_1 - n_1/\tau_r,$$

$$\frac{\partial n_2}{\partial t} = W_1 n_1 - \tilde{\alpha}n_2 - n_2/\tau_r.$$

The first conduction state, $n_0$ is filled with a $W_{\text{PI}}$ rate (MPI or TI depending on the intensity) obtained from the Keldysh theory[46]. This stage describes the primary photo-ionization process. Each conduction state is bridged through one-photon absorption similar to the mechanism of inverse Bremsstrahlung absorption through the rate $W_1 = 3.5 \times 10^{-7} E_L^2$ in units of s⁻¹, where $E_L$ is the laser electric field in units of V/m. The SSI model includes CI and possible electron avalanche ionization as highly energetic electrons in the conduction band may transfer a fraction of their energy by collisions with electrons in the valence band. The last state corresponds to the minimum energy required to induce impact ionization (i.e., at least 1.5 times the bandgap[34]) with the rate $\tilde{\alpha} = 10^{15}$ s⁻¹. Conduction electrons can also recombine within a characteristic time-scale of $\tau_r = 1$ ps. These parameter values were already used in various studies compatible with the present conditions[47–50]. The free electron density is $n_e = n_0 + n_1 + n_2$. Since the emptying of the valence band is not accounted for, $n_e$ can exceed tens of critical plasma densities. This approach remains valid as long as the band structure remains intact, i.e., before melting occurs.

### c. Two-temperature model (TTM)

Following Refs. [30,33] and references therein, we use a standard two-temperature model (TTM),

$$C_e \frac{\partial T_e}{\partial t} = \frac{\partial U}{\partial t} - \frac{3}{2} k_B \frac{\partial n_e}{\partial t} T_e - G(T_e - T_l),$$

$$C_l \frac{\partial T_l}{\partial t} = G(T_e - T_l).$$

The heat capacities are $C_e = 3n_e k_B/2$ and $C_l = 3n_a k_B/2$. The electron-ion energy exchange factor $G$ is evaluated by $G = C_e \nu_c m_e/m_a$. $n_a$ and $m_a$ are the carbon atomic density and mass, respectively. The electron-to-ion mass ratio weights the collision frequency to account for energy exchange ($\nu_c$ accounts for momentum transfer). The source term is evaluated with the Drude model:

$$\frac{\partial U}{\partial t} = \frac{e^2 n_e \nu_c}{m_e(\omega^2 + \nu_c^2)} E_L^2$$

where $\nu_c = \nu_{\text{ph}}$ is the electron-phonon collision frequency as the collisions are mainly driven by phonons in the solid state. It then reads $\nu_{\text{ph}} = \nu_{\text{ph0}} T_l/T_0$, and $\nu_{\text{ph0}}$ is the electron-phonon collision frequency at room temperature $T_0 = 300$ K. It is set to $\nu_{\text{ph0}} = 10^{14}$ s$^{-1}$ [48]. $\nu_c$ is limited to $5 \times 10^{15}$ s$^{-1}$ to account for the upper value of the collision frequency imposed by the electron mean free path[33].

### d. Two-step interaction model (TSI)

In our TSI model, the PIC simulation starts with a homogeneous plasma slab with a density $n_e^m \approx 70 n_c$ of the initial target thickness. The electrons and carbon ions species are initialized with Maxwell-Boltzmann distribution functions with temperatures $T_e^m$ and $T_i^m$, respectively, computed in the TTM. Since $n_e^m$ exceeds the carbon solid atomic density $n_a = 62 n_c$, the plasma is modeled partially ionized with a mixture of single and double ionization states of carbons $C^{1+}$ and $C^{2+}$ with $n_{C^{1+}} = 54 n_c$ and $n_{C^{2+}} = 8 n_c$, respectively. In the low-intensity range with $I < 10^{12}$ W/cm$^2$ where the SSI description holds, we expect full single ionization of carbon atoms by MPI reaching the density $n_{C^{1+}} = n_a$ before a significant number of $C^{2+}$ is produced. Collisions are of minor importance because electrons are only moderately heated by the laser in this intensity range. With increasing intensity, a fraction of these ions is further ionized to $C^{2+}$ to reach $n_e^m$. The other simulation parameters are kept the same as in the section PIC above.

### Acknowledgements

The research leading to these results has received funding from LASERLAB-EUROPE (Grant Agreement No. 871124, European Union's Horizon 2020 research and innovation program) and from the Bundesministerium für Bildung und Forschung (BMBF, Grants Agreement No. 03VNE2068D, No. 03Z1H531, No. 05K16SJC, No. 05K19SJC, No. 05P15SJFA1, and No. 05P19SJFA1).

### Author details
[1]Helmholtz Institute Jena, Fröbelstieg 3, 07743 Jena, Germany.
[2]Institute of Optics and Quantum Electronics, Friedrich-Schiller-Universität Jena, Max-Wien-Platz 1, 07743 Jena Germany.
[3]CEA-CESTA, 15 Avenue des Sablières, CS60001, 33116 Le Barp Cedex, France.
[4]Institut Lumière Matière, UMR 5306 - CNRS, Université de Lyon 1, 69622 Villeurbanne, France.
[5]LULI, CNRS, CEA, Sorbonne Université, Institut Polytechnique de Paris, Palaiseau, France.
[6]Université de Bordeaux-CNRS-CEA, CELIA, UMR 5107, Talence, France.

### Author contributions
Y. A., G. A. B., S. K., A. K., M. Ho., I. T., S. H., M. He. and F. S. setup and conducted the experiments.
Y. A., G. A. B., S. K. and M. C. K analyzed and interpreted the experimental data.
G. D., S. S., M. G., F. C. and Y. A. conducted the modeling developments and the numerical simulations.
G. D., S. S., Y. A. and M. C. K. analyzed and interpreted the simulation data.
Y. A. wrote the manuscript and all authors contributed to its improvement.
M. C. K and Y. A. led the project.

### Data availability
The experimental and simulation data that support the findings of this work are available from the corresponding author upon reasonable request.

### Conflict of interest
The authors declare no competing interests.